\documentclass[11pt,a4paper]{article}
\usepackage[margin=1in]{geometry} %margins
\usepackage{graphicx} %for pics
\graphicspath{{J:/4274/Medicaid idea/}}
\usepackage{caption}
\usepackage{listings}
\usepackage{xparse}
\usepackage{setspace}
\usepackage{amsmath}
\usepackage{blindtext}
\usepackage{apacite}
\usepackage{natbib}
\usepackage[graphicx]{realboxes}
\usepackage[toc,page]{appendix}
\usepackage{float}
\usepackage{threeparttable}
\usepackage{enumitem}
\usepackage{amsmath,amsfonts,amssymb,amsthm}
\usepackage{url}
\usepackage{longtable,tabu}

\usepackage{chngcntr}

\title{An Empirical Evaluation On The Effectiveness Of Medicaid Expansion Across 49 States \footnote{Including Washington DC. Montana and Louisiana have been omitted from the study as they expanded Medicaid during 2016 together with the lack of data for 2017.} \footnote{ECON 4274 Final Project, HKUST Spring Semester 2018}}
\date{May 25 2018}
\author{Chan Tung Yu Marco}

\spacing{1.25}

\begin{document}

\maketitle

\begin{abstract}
\normalsize

In 2014 the Patient Protection and Affordable Care Act (ACA) introduced the expansion of Medicaid where states can opt to expand the eligibility for those in need of free health insurance. In this paper we attempt to assess the effectiveness of Medicaid expansion on health outcomes of state populations using  Difference-in-Difference (DD) regressions to seek for causal impacts of expanding Medicaid on health outcomes in 49 states. We find that in the time frame of $2013$ to $2016$, Medicaid expansion seems to have had no significant impact on the health outcomes of states that have chosen to expand.  

\end{abstract}

\section{Introduction}
\indent\indent The Patient Protection and Affordable Care Act (ACA) more commonly known as Obamacare is one of the largest reformations of the U.S. healthcare system. One of the main features of the ACA is the revised expansion of Medicaid eligibility in 2014. The Medicaid program has been providing and funding medical assistances to American citizens since its introduction in 1965. As of 2018, it has been responsible for providing free health insurance to 68 million people \citep{Med}.
 The Medicaid Expansion under the ACA allows individuals and households to qualify for Medicaid based on income alone with respect to the Federal Poverty Level (FPL). Under the current law, individual states have the choice on whether or not to participate in the ACA.  Since its introduction, the ACA has sparked controversy within various fields ranging from politics, businesses and the general population, particularly people in the lower income groups. This is understandable from an economic viewpoint as it is easy to see the costs of providing free health insurance for millions. This brings us to the question of whether the costs of such policies can be justified by the benefits.

\indent To see the relevance of Medicaid, we look at the following observations. Over the years, it has been observed that healthcare costs have been experiencing a steady price inflation.\footnote{see Appendix \ref{appendix: health price}} \citet{wilsoneconomic} suggests that this is essentially a demand-pull inflation where he narrows down to two main causes. The first being that people are getting richer over time (as shown by the steady increase in real GDP per capita) which leads to an income effect of increased healthcare consumption. The second being that people are valuing personal health more and more \citep{spogard1999governance} hence leading to increased healthcare consumption.\footnote{see Appendix \ref{appendix: health consumption}} However, despite the increasing GDP per capita, the income gap has been observed to be widening since the 1980s.\footnote{see Appendix \ref{appendix: income gap}} This is wherein lies the relevance of Medicaid, to provide healthcare to the lower income groups who are gradually leaving the healthcare market due to inflating prices. 

\citet*{boudreaux2016long} finds that exposure to Medicaid in early childhood is associated with improvements in adult health which was specific to the treatment group having participated in Medicaid. \citet{thompson2017long} finds similar results "an additional year of public health insurance eligibility during childhood improves the summary index of adult health." So it should be safe to say that Medicaid has been shown to have a positive effect on health outcomes. From this, intuition would tell us that by extension, Medicaid expansion would also have a positive effect on health outcomes.    

As discussed by \citet{wilsoneconomic}, there are many impacts (both  budgetary and non-budgetary) brought about as a result. One of the most discussed non-budgetary concerns is the crowding out of the health insurance market, although according to \citet{yazici2000medicaid} "Medicaid expansions resulted in relatively little crowd out of private insurance." So in order to settle the controversy around policies such as Medicaid and by extension the ACA, government officials and the general public have to be convinced that the overall benefits of the program are significant enough to justify the costs (both budgetary and non-budgetary). In this paper, we will attempt to evaluate whether Medicaid expansion has a significant impact on health outcomes by seeing whether states that have chosen to expand Medicaid experience significant differences in health status. Because this topic is of great interest to a great number of parties, this paper attempts to shed light on the apparent effectiveness of the ACA's main objective of promoting health status of the states. We will do this by using Difference-in-Difference (DD) regressions to seek a causal relationship of having expanded Medicaid on health outcomes of the states.  

\section{Data}
\indent\indent The data used in this study come from a variety of sources, primarily from the U.S. Census Bureau, The State Health Access Data Assistance Center (SHADAC) and the Henry J. Kaiser Family Foundation (KFF). The structure of the dataset is in panel data. Most of the data (control variables) are in terms of either percentages or rates per $n$ people to control for population sizes. Most control variables are time-variant whilst two are time-invariant due to the lack of data.\footnote{The two time-invariant controls being Human Development Index (HDI) and CO$_2$ emissions.} For HDI, we assume that there is little to no fluctuation. This can be justified from the general observation that High Income Countries (HIC) tend to have stable measures of quality of life. This is shown by \citet{konya2008does} where they conclude that more developed countries tend to converge rather slowly with respect to HDI. As for CO$_2$ emissions, we take a fifteen year average from 2000 to 2015. 

The time periods of the dataset go from 2013 to 2016 where the data is annual. Some controls that are expressed as a percentage of population use annual estimates of population by the U.S. Census Bureau using 2010 as an estimate base.\footnote{Number of hospitals per 100,000, Death rate and Number of violent crimes uses the annual population estimates from the U.S. Census Bureau.} The controls are chosen with considerations of how important they are as determinants of health status. Almost all controls are proxies or measures of the Social Determinants of Health recognized by the World Health Organization (WHO), \citep{national2016healthy}.   

\begin{table}[h] 
\centering

\caption{Statistic Definitions}
\begin{threeparttable}
\begin{tabular}{c c c}

\hline\hline
Statistic & & Definition \\ 
\hline

treat.1 & & Dummy $(=1)$ for states that expanded Medicaid during 2014\tnote{1}\\
treat.2 & & Dummy $(=1)$ for states that expanded Medicaid during 2015\tnote{1} \\
D.14 & & Time dummy for 2014 \\
D.15 & & Time dummy for 2015 \\
D.16 & & Time dummy for 2016 \\
HDI & & Human Development index (HDI)\tnote{2} \\
co2 & & Mean CO$_2$ emissions from 2000-2015 in million metric tons\tnote{3} \\
rGDP.cap & & Real GDP per capita in chained 2009 dollars\tnote{4} \\
disability & & Percent of adults who report having a disability\tnote{5}\\
Y1 & & Health Outcome: Prevalence of Diabetes, CVD and Asthma in adults\tnote{6}  \\
Y2 & & Health Outcome: Percent of adults with fair or poor health status\tnote{6} \\
hos.days & & Hospital inpatient days per 1000 population\tnote{1} \\
pov.perc & & Percentage in poverty\tnote{5}~ \tnote{7} \\
HS.perc & & Percentage that have attained High School education or higher\tnote{5}  \\
uni.perc & & Percentage that have attained Bachelor's education or higher\tnote{5} \\
SSI.perc & & Percentage participating in the Supplemental Security Income program\tnote{8} \\
UE & & Unemployment rate\tnote{9} \\
age & & Median age\tnote{5} \\
obese & & Prevalence of obesity (BMI $\geq30$) among adults\tnote{6} \\
crime & & Violent crime rate per 100,000\tnote{10} \\
hosps & & Number of hospitals per 100,000\tnote{1} \\
insured & & Percentage of health insured\tnote{6} \\
exercise & & Percentage of adults participating in any physical activity or exercise\tnote{5} \\
death & & Health Outcome: Deaths per 100,000\tnote{5} \\

\hline
\end{tabular}

\begin{tablenotes}

\item[1] \scriptsize{\citet{KFF}}

\item[2] \scriptsize{\citet{hdi}}

\item[3] \scriptsize{\citet{co2}}

\item[4] \scriptsize{\citet{gdp}}

\item[5] \scriptsize{\citet{disability}; The definition of disability follows from the American Community Survey (ACS)}

\item[6] \scriptsize{\citet{shadac}}

\item[7] \scriptsize{As defined and measured by the U.S. Census Bureau}

\item[8] \scriptsize{\citet{bls}}

\item[9] \scriptsize{\citet{ssa}}

\item[10] \scriptsize{\citet{fbi}}

\end{tablenotes}
\end{threeparttable}
\label{table: Definitions}
\end{table}

~\\
\indent Table \ref{table: Definitions} lists all the variables and controls used in the study together with their definitions and Table \ref{summary statistics} shows the summary statistics. \citet{national2016healthy} lists a number of social determinants of health including economic stability, education, health and healthcare, neighbourhood and built environment. The variables listed in Table \ref{table: Definitions} attempt to control for the determination of these variables on health outcomes.

\begin{table}[] \centering 
  \caption{Summary Statistics} 
  \label{summary statistics}  
\begin{tabular}{@{\extracolsep{5pt}}lccccc} 
\\[-1.8ex]\hline 
\hline \\[-1.8ex] 
Statistic & \multicolumn{1}{c}{N} & \multicolumn{1}{c}{Mean} & \multicolumn{1}{c}{St. Dev.} & \multicolumn{1}{c}{Min} & \multicolumn{1}{c}{Max} \\ 
\hline \\[-1.8ex] 
treat.1 & 196 & 0.531 & 0.500 & 0 & 1 \\ 
treat.2 & 196 & 0.061 & 0.240 & 0 & 1 \\ 
D.14 & 196 & 0.250 & 0.434 & 0 & 1 \\ 
D.15 & 196 & 0.250 & 0.434 & 0 & 1 \\ 
D.16 & 196 & 0.250 & 0.434 & 0 & 1 \\ 
HDI & 196 & 6.794 & 1.546 & 2.900 & 9.600 \\ 
co2 & 196 & 91.017 & 39.027 & 37.101 & 283.607 \\ 
rGDP.cap & 196 & 50,494.340 & 18,072.080 & 31,635 & 159,530 \\ 
disability & 196 & 0.131 & 0.022 & 0.095 & 0.202 \\ 
Y1 & 196 & 0.231 & 0.026 & 0.180 & 0.319 \\ 
Y2 & 196 & 0.144 & 0.033 & 0.090 & 0.273 \\ 
hos.days & 196 & 600.730 & 186.877 & 336 & 1,464 \\ 
pov.perc & 196 & 0.133 & 0.036 & 0.055 & 0.258 \\ 
HS.perc & 196 & 0.881 & 0.030 & 0.812 & 0.926 \\ 
uni.perc & 196 & 0.294 & 0.060 & 0.183 & 0.554 \\ 
SSI.perc & 196 & 0.024 & 0.008 & 0.010 & 0.043 \\ 
UE & 196 & 5.571 & 1.488 & 2.700 & 9.600 \\ 
age & 196 & 37.792 & 2.399 & 29.600 & 44.000 \\ 
obese & 196 & 0.292 & 0.035 & 0.202 & 0.377 \\ 
crime & 196 & 374.424 & 184.164 & 99.300 & 1,300.300 \\ 
hosps & 196 & 2.071 & 1.219 & 0.735 & 6.409 \\ 
insured & 196 & 0.900 & 0.040 & 0.781 & 0.975 \\ 
exercise & 196 & 0.755 & 0.043 & 0.619 & 0.843 \\ 
death & 196 & 8.426 & 1.224 & 5.129 & 12.293 \\ 
\hline \\[-1.8ex] 
\end{tabular} 
\end{table}

\section{Empirical Analysis}
\subsection{Empirical Strategy}
\indent\indent To attempt to find a causal relationship between health outcomes of the states and Medicaid expansion status, the potential endogeneity surrounding whether a state chooses to expand Medicaid or not should be addressed as this makes for essentially non-random treatment. As mentioned earlier, to tackle this endogeneity problem we will be using DD regression. In addition, among the 49 states in our study, most states which have expanded Medicaid did so during the year 2014 whilst three states expanded during 2015 and two states expanded during 2016 (which have been omitted in this study). To attempt capture the effect of Medicaid expansion across all 49 states, we will be using two treatment periods in our DD regression (2014 and 2015). For the DD estimators to be valid in the face of endogeneity we need to make two identification assumptions. 

\newtheorem{assumption}{Assumption}

\begin{assumption}[No anticipation]
\label{no anticipation}
Given a treatment group of individuals $i=1,2,..,n$ and control group of individuals $j=1,2,...,m$ and time periods $t = 0,1$ with outcome $Y$, and treatment dummy $T$,
\[
\mathrm{E}(Y_i|t=0,T=1)=\mathrm{E}(Y_i|t=0,T=0)
\label{no anticipation eq}
\tag{1}
\] 
Meaning that the outcome of the treatment group at $t=0$ is independent of their status of being in the treatment group. 
\end{assumption}

\begin{assumption}[Parallel Trend] 
\label{parallel trend}
The treatment group would have followed the same trend as the control group given that the treatment was absent.
\\ \indent i.e.
\[
\mathrm{E}(Y_i|t=1,T=0)-\mathrm{E}(Y_i|t=0,T=0)=\mathrm{E}(Y_j|t=1,T=0)-\mathrm{E}(Y_j|t=0,T=0)
\label{parallel trend eq}
\tag{2}
\]

\end{assumption}

Since the treatment in our study has an indirect effect on health status in the sense that it is a vehicle to health care (which has a direct impact on health) access through Medicaid, the validity of assumption \ref{no anticipation} is transferred over to the anticipation of healthcare on health outcomes. In other words, an individual's anticipation of being in a state with Medicaid expansion should not affect his/her health outcome directly. Medicaid expansion is merely a vehicle to healthcare access which itself is a vehicle to healthcare which has a direct impact on health outcomes. Therefore any anticipation of enjoying Medicaid expansion or not should not directly affect an individual's health outcome.

For assumption \ref{parallel trend}, we need to show that both groups would have followed the same trend given neither were treated. To test this assumption, we ran a Welch two sample t-test to test if there were significant differences between $\Delta_{T_1=1}=\mathrm{E}(Y_i,t=2014,T_1=1)-\mathrm{E}(Y_i,t=2013,T_1=1)$ and $\Delta_{T_1=0}=\mathrm{E}(Y_j,t=2014,T_1=0)-\mathrm{E}(Y_j,t=2013,T_1=0)$.\footnote{In this case $T_1=0$ implies control group.} Although this is not a perfect test in that despite during the year 2013 none of the groups were treated, the first treatment group expanded Medicaid during 2014, meaning that the difference between health outcomes from 2013 and 2014 is not entirely the case where both groups were untreated (did not enjoy Medicaid expansion). However it could also be argued that the benefits of Medicaid expansion on health outcomes are not immediately observable.

After performing the Welch t-test on all three health outcomes, we observe that for $Y_1$ the $p$-value $=0.2761$ and for $death$ $p$-value$=0.8477$ whilst for $Y_2$ the $p$-value $=0.003114$. We can see from this that the validity of assumption \ref{parallel trend} can be somewhat justified as the $p$-values for $Y_1$ and $death$ allow us to accept the null of the health trends being statistically similar. However we should be wary of the output of the DD regression on $Y_2$ as assumption \ref{parallel trend eq} does not seem to hold in this case. 

Other than the potential endogeneity surrounding whether a state chooses to expand or not, we will assume for our study that all controls are exogenous for the sake of simplification.     

\subsection{Econometric Model}
The DD regression model will take the following form.

$$Health=\alpha+\delta_1 D_{14}+\delta_2 D_{15}+\delta_3 D_{16}+\delta_4 T_1+\delta_5(D_{14}\times T_1)+\delta_6(D_{15}\times T_1)+\delta_7(D_{16}\times T_1)$$
\[
+ \delta_8 T_2+\delta_9(D_{14}\times T_2)+\delta_{10}(D_{15}\times T_2)+\delta_{11}(D_{16}\times T_2)+X\beta+\epsilon
\tag{3}
\label{DD general form}
\]

Where $X$ is the matrix of controls. \par

To interpret the results from the DD regression, we will be comparing the health outcomes of the control group with states that expanded during 2014 and states that expanded during 2015 at the year 2016. This change would be represented by $\delta_4+\delta_7$ for states that expanded during 2014 and for states that expanded during 2015, $\delta_8+\delta_{11}$.

As listed in Table \ref{table: Definitions} we have three measures of health outcomes. We will run DD regressions on all three outcomes and compare the results. 

\subsection{Empirical Results}
\subsubsection{$Y_1$: Prevalence of Diabetes, Cardiovascular Diseases and Asthma in Adults}
\indent\indent Table \ref{Y1} shows the DD regressions for the health outcome $Y_1$ with heteroskedastic robust standard errors. We can see that none of the DD coefficients are significant at any level and are also very small in magnitude as well as occasionally having non-intuitive signs. The controls across the six specifications are shown to be very significant as well as most of them having signs that are consistent with intuition.\footnote{Though for some controls such as hos.days, there may be some reverse causality that explains their signs (having CVD, Asthma or diabetes would lead to one staying at the hospital longer.)} Further to the right of Table \ref{Y1}, controls that have been shown to be insignificant are dropped and those significant are kept. We see that for the controls that are kept that their coefficients are relatively stable across specifications. The $R^2$ values are also very high for specifications (2) to (6) and the $F$-statistics are shown to be highly significant for all specifications. 

After reviewing the results from Table \ref{Y1} it seems apparent that expanding Medicaid in either 2014 or 2015 has not shown any significant changes in the prevalence of Diabetes, CVD and Asthma in Adults after having incorporated controls derived from the social determinants of health which are shown to be highly significant and relevant. 

\subsubsection{$Y_2$: Percentage of Adults with fair or poor health status}
\indent\indent Looking at Table \ref{Y2}, we actually see $(D_{14} \times T_1)$ to be significant at the $5\%$ level but with a positive sign for specifications (4), (5) and (6) whilst all the other DD coefficients are insignificant as with $Y_1$. The controls also seem relatively stable, although $HS.perc$ changes from specification (3) to (4) but stays stable up to (6). The $R^2$ values are relatively high as you go towards the right of Table \ref{Y2} with specification (6) with $\bar{R^2}=0.788$. The $F$-statistics are also shown to be highly significant for all specifications. 

As mentioned in section 3.1, we could not validate assumption \ref{parallel trend eq} for $Y_2$. Reviewing the results in table \ref{Y2}, Medicaid expansion seems to have had an adverse effect on the reported health status of adults. Although this result seems counter-intuitive, it is possible that this is a result of an omitted factor from our model. There is also the possibility of reverse causality, but further study is required to affirm whether Medicaid expansion does lead to higher counts of fair or poor health status amongst adults. This may of be some interest as positive signs on the DD coefficients are observed throughout all specifications for $(D_{14} \times T_1)$, $(D_{15} \times T_1)$, $(D_{16} \times T_1)$, $(D_{14} \times T_2)$ and $(D_{16} \times T_2)$. 

\subsubsection{Death Rate}
\indent\indent The use of death rate as a measure of health status is commonly discouraged as there are a multitude of factors that go into the determination of the figure. Because of this, it is very easy to run into omitted variable bias (OVB) and endogeneity problems in estimation when using death rate as an outcome. In this paper, we will be running two sets of regressions on death rate. In the U.S., over $50\%$ of deaths are caused by heard diseases \citep{cdcheart}. As such, we will be running one set of regressions using $Y_1$    as a control and in another we will omit $Y_1$ from the regressions. By including $Y_1$ in the regression it is apparent that the model will suffer from some form of endogeneity as we have been using it as an outcome variable itself; and therefore lead to the estimates to be biased and inconsistent. However in exchange, the regression models should have a much better fit with the data due to the correlation and relevance of $Y_1$ on $death$. 

For the second set of regressions, the presence of the problem of inconsistent and biased estimates depend on whether $Y_1$ is correlated with treatment. If $Y_1$ is correlated with our variables of interest then by omitting it we run into OVB. However if there is no correlation then the estimates will be free of bias (from lack of correlation) and inconsistency (from omission of endogenous control). Table \ref{biserial} shows the Point-Biserial correlation between $death$ and the DD variables. We can see that all of them have relatively low values of $r_{pb}$, so it is possible that by omitting $Y_1$, the DD coefficients will be unbiased and consistent. 

\begin{table}[ht]
\centering
\caption{Biserial correlation with $death$}
\begin{tabular}{c c c}
\hline
DD variable & $r_{pb}$ \\
\hline
$T_1$ & $-5.2\times 10^{-7}$\\ 
$T_2$ & $0.0324$\\
$D_{14}\times T_1$ & $0.0232$\\
$D_{14}\times T_2$ & $0.0216$\\
$D_{15}\times T_1$ & $0.0115$\\
$D_{15}\times T_2$ & $0.0193$\\
$D_{16}\times T_1$ & $-0.0957$\\
$D_{16}\times T_2$ & $-0.0133$\\

\hline
\end{tabular}
\label{biserial}
\end{table}

\newpage
We can see from table \ref{death1} that $Y_1$ is highly significant across all specifications as well as having coefficients of high magnitude. We see some significance for the time dummy coefficients for 2014 and 2015. Looking at specifications (3), (4) and (5) we find that $(D_{14}\times T_1)$ is shown to be significant as well as having a negative sign. Specification (3) also shows $(D_{14}\times T_1)$, $(D_{15}\times T_1)$ and $(D_{16}\times T_1)$ to be significant at the $10\%$ level. 
The highly significant controls also seem to have reasonable signs on their coefficients with the exception of $insured$ which may be subject to reverse causality\footnote{Residents of states with high death rates may tend to buy insurance more than states with lower death rates.}. We also see very high goodness of fits with all five specifications having $\bar{R^2}$ values between $0.8$ and $0.9$ together with high joint significance shown by the $F$-statistics.  

Although the results from table \ref{death1} seem very intuitive, it should be noted that these estimates are known suffer from endogeneity meaning we cannot draw any causal conclusions. At best, the specifications in table \ref{death1} show the correlation between the covariates.

Table \ref{death2} shows the second set of regressions where we omit $Y_1$ as a control. With these specifications we are more convinced the estimates are free of bias and inconsistency. Similar with previous cases, none of the DD coefficients are shown to be significant at any level, although the signs are more intuitive this time. The controls are also relatively stable and unchanging  which convinces us that specification (6) may be accurate. In theory, by including $Y_1$ as a control in the regressions in table \ref{death1}, we run the trade-off of increasing goodness of fit at the cost of consistent and unbiased estimates when compared to the regressions on table \ref{death2}. However, the $\bar{R^2}$ values are surprisingly high and not so different from the regressions from table \ref{death1}.       

\section{Conclusion}
\indent\indent After reviewing the results of the regressions on all three health outcomes: Prevalence of Diabetes, CVD and Asthma $(Y_1)$, adults with fair or poor health status $(Y_2)$ and Death rate it seems apparent that Medicaid expansion by the ACA has not had any significant effect on the mentioned health outcomes in the time frame of $2013$ to $2016$. Although the set of regressions for death rate that included $Y_1$ as a control showed intuitive and reasonable results for states that expanded Medicaid in 2014, the concerned estimates suffer from endogeneity and no causal interpretation can be drawn.

It should be noted that this study may not be the best reflection on the effectiveness of Medicaid Expansion on health outcomes as the time frame of this study is relatively short when compared to other studies that have attempted to study the effectiveness of Medicaid.\footnote{The studies provided by \citet{boudreaux2016long}, \citet{thompson2017long}, \citet{sohn2017medicaid} and many others who have studied the effectiveness of Medicaid have studied a much larger time frame.} This may suggest that if Medicaid expansion does indeed have a significant impact on health outcomes, that the effects are not immediately observable and require a study of a longer time frame. In terms of a short-term evaluation on the effectiveness of Medicaid expansion, it may be more effective to evaluate the impact on healthcare usage as to health outcomes, as healthcare usage is more immediately observable as compared to health outcomes. Thus, in order to have a better grasp of this issue this study may have to be revisited in a few years time.   

\clearpage

\begin{longtable}{@{\extracolsep{0.1pt}}lcccccc} 
\caption{$Y_1$: Prevalence of Diabetes, CVD and Asthma in Adults} 
  \label{Y1} 
\small 
\\[-1.8ex]\hline 
\hline \\[-1.8ex] 
 & \multicolumn{6}{c}{\textit{Dependent variable:}} \\ 
\cline{2-7} 
\\[-1.8ex] & \multicolumn{6}{c}{Y1} \\ 
\\[-1.8ex] & (1) & (2) & (3) & (4) & (5) & (6)\\ 
\hline \\[-1.8ex] 
 D.14 & $-$0.001 & $-$0.001 & $-$0.002 & $-$0.001 & 0.002 & 0.001 \\ 
  & (0.005) & (0.004) & (0.003) & (0.003) & (0.003) & (0.003) \\ 
  & & & & & & \\ 
 D.15 & $-$0.005 & $-$0.003 & $-$0.004 & $-$0.003 & $-$0.002 & $-$0.006$^{*}$ \\ 
  & (0.005) & (0.004) & (0.003) & (0.003) & (0.003) & (0.003) \\ 
  & & & & & & \\ 
 D.16 & $-$0.001 & 0.001 & $-$0.001 & 0.001 & 0.003 & $-$0.0002 \\ 
  & (0.005) & (0.004) & (0.003) & (0.003) & (0.003) & (0.003) \\ 
  & & & & & & \\ 
 treat.1 & $-$0.0001 & 0.002 & 0.0001 & 0.002 & 0.002 & 0.001 \\ 
  & (0.005) & (0.004) & (0.003) & (0.003) & (0.003) & (0.003) \\ 
  & & & & & & \\ 
 treat.2 & $-$0.006 & $-$0.0004 & $-$0.004 & $-$0.0002 & $-$0.001 & $-$0.001 \\ 
  & (0.007) & (0.007) & (0.005) & (0.004) & (0.003) & (0.003) \\ 
  & & & & & & \\ 
 obese & 0.446$^{***}$ & 0.163$^{***}$ & 0.277$^{***}$ & 0.223$^{***}$ & 0.183$^{***}$ & 0.180$^{***}$ \\ 
  & (0.044) & (0.040) & (0.034) & (0.032) & (0.039) & (0.036) \\ 
  & & & & & & \\ 
 age & 0.005$^{***}$ & 0.002$^{***}$ & 0.002$^{***}$ & 0.001$^{***}$ & 0.002$^{***}$ & 0.001$^{***}$ \\ 
  & (0.0005) & (0.0005) & (0.0004) & (0.0004) & (0.0004) & (0.0004) \\ 
  & & & & & & \\ 
 disability &  & 0.766$^{***}$ & 0.752$^{***}$ & 0.678$^{***}$ & 0.695$^{***}$ & 0.728$^{***}$ \\ 
  &  & (0.064) & (0.057) & (0.055) & (0.055) & (0.055) \\ 
  & & & & & & \\ 
 rGDP.cap &  &  & 0.00000$^{***}$ & $-$0.00000$^{**}$ &  &  \\ 
  &  &  & (0.00000) & (0.00000) &  &  \\ 
  & & & & & & \\ 
 hosps &  &  & $-$0.005$^{***}$ & $-$0.008$^{***}$ & $-$0.006$^{***}$ & $-$0.006$^{***}$ \\ 
  &  &  & (0.001) & (0.001) & (0.001) & (0.001) \\ 
  & & & & & & \\ 
 hos.days &  &  &  & 0.00004$^{***}$ & 0.00003$^{***}$ & 0.00002$^{***}$ \\ 
  &  &  &  & (0.00001) & (0.00000) & (0.00001) \\ 
  & & & & & & \\ 
 exercise &  &  &  &  & $-$0.078$^{**}$ & $-$0.101$^{***}$ \\ 
  &  &  &  &  & (0.031) & (0.032) \\ 
  & & & & & & \\ 
 insured &  &  &  &  &  & 0.098$^{***}$ \\ 
  &  &  &  &  &  & (0.031) \\ 
  & & & & & & \\ 
 D.14:treat.1 & 0.002 & 0.003 & 0.002 & 0.003 & 0.002 & 0.0004 \\ 
  & (0.007) & (0.005) & (0.005) & (0.004) & (0.004) & (0.004) \\ 
  & & & & & & \\ 
 D.15:treat.1 & 0.002 & 0.001 & 0.001 & 0.001 & 0.0002 & $-$0.002 \\ 
  & (0.007) & (0.005) & (0.005) & (0.004) & (0.004) & (0.004) \\ 
  & & & & & & \\ 
 D.16:treat.1 & 0.002 & 0.002 & 0.002 & 0.003 & 0.002 & $-$0.0001 \\ 
  & (0.007) & (0.005) & (0.005) & (0.005) & (0.005) & (0.004) \\ 
  & & & & & & \\ 
 D.14:treat.2 & $-$0.001 & $-$0.001 & $-$0.001 & $-$0.001 & $-$0.0005 & 0.0001 \\ 
  & (0.013) & (0.013) & (0.011) & (0.008) & (0.008) & (0.007) \\ 
  & & & & & & \\ 
 D.15:treat.2 & 0.011 & 0.007 & 0.008 & 0.008 & 0.007 & 0.006 \\ 
  & (0.012) & (0.011) & (0.010) & (0.008) & (0.007) & (0.006) \\ 
  & & & & & & \\ 
 D.16:treat.2 & 0.001 & $-$0.001 & $-$0.001 & $-$0.001 & $-$0.001 & $-$0.001 \\ 
  & (0.015) & (0.014) & (0.013) & (0.010) & (0.010) & (0.008) \\ 
  & & & & & & \\ 
 Constant & $-$0.080$^{***}$ & 0.001 & $-$0.028 & 0.019 & 0.074$^{**}$ & 0.030 \\ 
  & (0.023) & (0.018) & (0.017) & (0.017) & (0.034) & (0.034) \\ 
  & & & & & & \\ 
\hline \\[-1.8ex] 
Observations & 196 & 196 & 196 & 196 & 196 & 196 \\ 
R$^{2}$ & 0.590 & 0.770 & 0.826 & 0.856 & 0.858 & 0.865 \\ 
Adjusted R$^{2}$ & 0.561 & 0.752 & 0.811 & 0.843 & 0.844 & 0.852 \\ 
Residual Std. Error & 0.017 & 0.013 & 0.011 & 0.010 & 0.010 & 0.010 \\ 
F Statistic & 20.173$^{***}$ & 43.220$^{***}$ & 53.256$^{***}$ & 62.479$^{***}$ & 63.196$^{***}$ & 63.182$^{***}$ \\ 
\hline 
\hline \\[-1.8ex] 
\textit{Note:}  & \multicolumn{6}{r}{$^{*}$p$<$0.1; $^{**}$p$<$0.05; $^{***}$p$<$0.01} \\ 
\end{longtable}

\newpage

\begin{longtable}{@{\extracolsep{0.1pt}}lcccccc} 
\caption{$Y_2$: Percentage of adults with fair or poor health status} 
  \label{Y2} 
\small 
\\[-1.8ex]\hline 
\hline \\[-1.8ex] 
 & \multicolumn{6}{c}{\textit{Dependent variable:}} \\ 
\cline{2-7} 
\\[-1.8ex] & \multicolumn{6}{c}{Y2} \\ 
\\[-1.8ex] & (1) & (2) & (3) & (4) & (5) & (6)\\ 
\hline \\[-1.8ex] 
 D.14 & $-$0.008 & $-$0.005 & $-$0.012$^{***}$ & $-$0.004 & $-$0.003 & $-$0.004 \\ 
  & (0.007) & (0.005) & (0.005) & (0.005) & (0.004) & (0.005) \\ 
  & & & & & & \\ 
 D.15 & 0.005 & 0.009 & $-$0.004 & $-$0.0005 & 0.003 & $-$0.0001 \\ 
  & (0.008) & (0.007) & (0.005) & (0.005) & (0.005) & (0.005) \\ 
  & & & & & & \\ 
 D.16 & 0.004 & 0.010 & $-$0.007 & 0.001 & 0.005 & 0.001 \\ 
  & (0.007) & (0.007) & (0.005) & (0.005) & (0.005) & (0.005) \\ 
  & & & & & & \\ 
 treat.1 & $-$0.005 & $-$0.009 & $-$0.005 & $-$0.002 & $-$0.003 & $-$0.002 \\ 
  & (0.009) & (0.007) & (0.006) & (0.005) & (0.004) & (0.004) \\ 
  & & & & & & \\ 
 treat.2 & 0.014 & 0.005 & 0.011$^{**}$ & 0.008 & 0.007 & 0.008 \\ 
  & (0.009) & (0.006) & (0.005) & (0.005) & (0.006) & (0.005) \\ 
  & & & & & & \\ 
 rGDP.cap & $-$0.00000$^{***}$ &  &  &  &  &  \\ 
  & (0.00000) &  &  &  &  &  \\ 
  & & & & & & \\ 
 pov.perc & 0.504$^{***}$ & 0.199$^{**}$ & $-$0.057 &  &  &  \\ 
  & (0.085) & (0.081) & (0.069) &  &  &  \\ 
  & & & & & & \\ 
 co2 & 0.00000 &  &  &  &  &  \\ 
  & (0.0001) &  &  &  &  &  \\ 
  & & & & & & \\ 
 HDI & 0.003$^{**}$ &  &  &  &  &  \\ 
  & (0.001) &  &  &  &  &  \\ 
  & & & & & & \\ 
 age &  & 0.004$^{***}$ & $-$0.001$^{**}$ & $-$0.001$^{***}$ & $-$0.002$^{***}$ & $-$0.001$^{***}$ \\ 
  &  & (0.001) & (0.001) & (0.001) & (0.001) & (0.001) \\ 
  & & & & & & \\ 
 UE &  & 0.004$^{*}$ & $-$0.0002 & 0.001 & 0.003$^{**}$ & 0.001 \\ 
  &  & (0.002) & (0.001) & (0.001) & (0.001) & (0.001) \\ 
  & & & & & & \\ 
 hosps &  & 0.003 &  &  &  &  \\ 
  &  & (0.002) &  &  &  &  \\ 
  & & & & & & \\ 
 disability &  &  & 1.133$^{***}$ & 0.969$^{***}$ & 0.995$^{***}$ & 0.993$^{***}$ \\ 
  &  &  & (0.096) & (0.085) & (0.082) & (0.082) \\ 
  & & & & & & \\ 
 HS.perc &  & $-$0.349$^{***}$ & $-$0.394$^{***}$ & $-$0.177$^{**}$ & $-$0.184$^{***}$ & $-$0.194$^{***}$ \\ 
  &  & (0.086) & (0.059) & (0.073) & (0.062) & (0.065) \\ 
  & & & & & & \\ 
 uni.perc &  & $-$0.207$^{***}$ &  &  &  &  \\ 
  &  & (0.032) &  &  &  &  \\ 
  & & & & & & \\ 
 SSI.perc &  &  &  & 0.159 &  &  \\ 
  &  &  &  & (0.247) &  &  \\ 
  & & & & & & \\ 
 exercise &  &  &  & $-$0.191$^{***}$ & $-$0.192$^{***}$ & $-$0.191$^{***}$ \\ 
  &  &  &  & (0.043) & (0.042) & (0.042) \\ 
  & & & & & & \\ 
 crime &  &  &  &  & $-$0.00002$^{***}$ &  \\ 
  &  &  &  &  & (0.00001) &  \\ 
  & & & & & & \\ 
 D.14:treat.1 & 0.010 & 0.013 & 0.014$^{**}$ & 0.012$^{**}$ & 0.012$^{**}$ & 0.012$^{**}$ \\ 
  & (0.010) & (0.008) & (0.007) & (0.006) & (0.006) & (0.006) \\ 
  & & & & & & \\ 
 D.15:treat.1 & 0.002 & 0.005 & 0.005 & 0.003 & 0.004 & 0.003 \\ 
  & (0.012) & (0.009) & (0.008) & (0.007) & (0.007) & (0.007) \\ 
  & & & & & & \\ 
 D.16:treat.1 & 0.004 & 0.005 & 0.004 & 0.002 & 0.003 & 0.003 \\ 
  & (0.010) & (0.008) & (0.007) & (0.006) & (0.006) & (0.006) \\ 
  & & & & & & \\ 
 D.14:treat.2 & 0.009 & 0.014 & 0.015 & 0.015 & 0.015 & 0.015 \\ 
  & (0.010) & (0.010) & (0.010) & (0.010) & (0.010) & (0.010) \\ 
  & & & & & & \\ 
 D.15:treat.2 & $-$0.011 & $-$0.007 & $-$0.009 & $-$0.009 & $-$0.009 & $-$0.009 \\ 
  & (0.011) & (0.008) & (0.007) & (0.007) & (0.008) & (0.007) \\ 
  & & & & & & \\ 
 D.16:treat.2 & $-$0.004 & 0.001 & $-$0.001 & 0.0004 & 0.001 & 0.0004 \\ 
  & (0.012) & (0.008) & (0.005) & (0.006) & (0.006) & (0.006) \\ 
  & & & & & & \\ 
 Constant & 0.096$^{***}$ & 0.319$^{***}$ & 0.401$^{***}$ & 0.362$^{***}$ & 0.383$^{***}$ & 0.374$^{***}$ \\ 
  & (0.012) & (0.080) & (0.063) & (0.059) & (0.052) & (0.053) \\ 
  & & & & & & \\ 
\hline \\[-1.8ex] 
Observations & 196 & 196 & 196 & 196 & 196 & 196 \\ 
R$^{2}$ & 0.550 & 0.687 & 0.782 & 0.806 & 0.814 & 0.806 \\ 
Adjusted R$^{2}$ & 0.513 & 0.657 & 0.762 & 0.788 & 0.796 & 0.788 \\ 
Residual Std. Error & 0.023 & 0.019 & 0.016 & 0.015 & 0.015 & 0.015 \\ 
F Statistic & 14.683$^{***}$ & 22.947$^{***}$ & 40.106$^{***}$ & 43.578$^{***}$ & 45.764$^{***}$ & 46.399$^{***}$ \\ 
\hline 
\hline \\[-1.8ex] 
\textit{Note:}  & \multicolumn{6}{r}{$^{*}$p$<$0.1; $^{**}$p$<$0.05; $^{***}$p$<$0.01} \\ 
\end{longtable} 

~\\[1cm]

\begin{longtable}{@{\extracolsep{0.1pt}}lccccc}
\caption{Death per 100,100 (with $Y_1$ control)} 
  \label{death1} 
\small  
\\[-1.8ex]\hline 
\hline \\[-1.8ex] 
 & \multicolumn{5}{c}{\textit{Dependent variable:}} \\ 
\cline{2-6} 
\\[-1.8ex] & \multicolumn{5}{c}{death} \\ 
\\[-1.8ex] & (1) & (2) & (3) & (4) & (5)\\ 
\hline \\[-1.8ex] 
 D.14 & 0.073 & 0.333$^{***}$ & 0.256$^{***}$ & 0.265$^{***}$ & 0.201$^{*}$ \\ 
  & (0.117) & (0.123) & (0.098) & (0.099) & (0.107) \\ 
  & & & & & \\ 
 D.15 & 0.094 & 0.150 & $-$0.113 & $-$0.032 & $-$0.095 \\ 
  & (0.124) & (0.119) & (0.114) & (0.105) & (0.115) \\ 
  & & & & & \\ 
 D.16 & 0.244$^{*}$ & 0.512$^{***}$ & 0.273$^{**}$ & 0.329$^{***}$ & 0.241$^{*}$ \\ 
  & (0.135) & (0.145) & (0.125) & (0.116) & (0.132) \\ 
  & & & & & \\ 
 treat.1 & $-$0.132 & $-$0.017 & $-$0.063 & 0.020 & 0.013 \\ 
  & (0.139) & (0.117) & (0.105) & (0.099) & (0.101) \\ 
  & & & & & \\ 
 treat.2 & $-$0.042 & $-$0.025 & 0.041 & 0.087 & 0.027 \\ 
  & (0.441) & (0.455) & (0.366) & (0.338) & (0.340) \\ 
  & & & & & \\ 
 Y1 & 24.193$^{***}$ & 15.217$^{***}$ & 8.438$^{***}$ & 9.924$^{***}$ & 8.326$^{***}$ \\ 
  & (2.674) & (2.795) & (2.582) & (2.319) & (2.564) \\ 
  & & & & & \\ 
 age & 0.191$^{***}$ & 0.219$^{***}$ & 0.195$^{***}$ & 0.185$^{***}$ & 0.202$^{***}$ \\ 
  & (0.020) & (0.016) & (0.017) & (0.016) & (0.017) \\ 
  & & & & & \\ 
 disability & 12.467$^{***}$ & 13.156$^{***}$ & 19.341$^{***}$ & 20.053$^{***}$ & 17.507$^{***}$ \\ 
  & (3.063) & (2.697) & (2.940) & (2.642) & (2.927) \\ 
  & & & & & \\ 
 co2 & 0.003$^{***}$ & 0.004$^{***}$ & 0.003$^{***}$ &  &  \\ 
  & (0.001) & (0.001) & (0.001) &  &  \\ 
  & & & & & \\ 
 HDI & $-$0.087$^{***}$ & $-$0.097$^{***}$ & $-$0.031 &  &  \\ 
  & (0.031) & (0.024) & (0.027) &  &  \\ 
  & & & & & \\ 
 exercise &  & $-$7.183$^{***}$ & $-$9.515$^{***}$ & $-$8.039$^{***}$ & $-$6.736$^{***}$ \\ 
  &  & (1.068) & (0.978) & (0.875) & (1.178) \\ 
  & & & & & \\ 
 pov.perc &  &  & 4.876$^{***}$ & 3.797$^{***}$ & 3.829$^{***}$ \\ 
  &  &  & (1.376) & (1.227) & (1.319) \\ 
  & & & & & \\ 
 insured &  &  & 9.783$^{***}$ & 6.874$^{***}$ & 7.165$^{***}$ \\ 
  &  &  & (1.429) & (1.364) & (1.725) \\ 
  & & & & & \\ 
 SSI.perc &  &  & $-$15.747$^{**}$ & $-$20.739$^{***}$ & $-$15.729$^{**}$ \\ 
  &  &  & (6.376) & (4.943) & (6.400) \\ 
  & & & & & \\ 
 hos.days &  &  &  & 0.001$^{***}$ & 0.001$^{***}$ \\ 
  &  &  &  & (0.0001) & (0.0002) \\ 
  & & & & & \\ 
 obese &  &  &  &  & 3.730$^{***}$ \\ 
  &  &  &  &  & (1.348) \\ 
  & & & & & \\ 
 crime &  &  &  &  & 0.0002 \\ 
  &  &  &  &  & (0.0002) \\ 
  & & & & & \\ 
 HS.perc &  &  &  &  & 0.113 \\ 
  &  &  &  &  & (2.148) \\ 
  & & & & & \\ 
 D.14:treat.1 & $-$0.085 & $-$0.121 & $-$0.285$^{**}$ & $-$0.233$^{*}$ & $-$0.230$^{*}$ \\ 
  & (0.181) & (0.161) & (0.145) & (0.139) & (0.138) \\ 
  & & & & & \\ 
 D.15:treat.1 & 0.019 & $-$0.050 & $-$0.259$^{*}$ & $-$0.188 & $-$0.156 \\ 
  & (0.192) & (0.165) & (0.147) & (0.140) & (0.141) \\ 
  & & & & & \\ 
 D.16:treat.1 & $-$0.048 & $-$0.107 & $-$0.283$^{*}$ & $-$0.214 & $-$0.203 \\ 
  & (0.201) & (0.180) & (0.155) & (0.146) & (0.145) \\ 
  & & & & & \\ 
 D.14:treat.2 & 0.020 & 0.032 & 0.032 & 0.026 & 0.014 \\ 
  & (0.523) & (0.563) & (0.444) & (0.415) & (0.428) \\ 
  & & & & & \\ 
 D.15:treat.2 & $-$0.159 & $-$0.141 & $-$0.180 & $-$0.157 & $-$0.109 \\ 
  & (0.591) & (0.587) & (0.432) & (0.413) & (0.428) \\ 
  & & & & & \\ 
 D.16:treat.2 & $-$0.039 & 0.003 & $-$0.085 & $-$0.061 & $-$0.073 \\ 
  & (0.554) & (0.595) & (0.491) & (0.457) & (0.487) \\ 
  & & & & & \\ 
 Constant & $-$5.734$^{***}$ & 0.430 & $-$5.332$^{***}$ & $-$4.327$^{***}$ & $-$6.734$^{***}$ \\ 
  & (0.601) & (1.203) & (1.205) & (1.109) & (1.390) \\ 
  & & & & & \\ 
\hline \\[-1.8ex] 
Observations & 196 & 196 & 196 & 196 & 196 \\ 
R$^{2}$ & 0.859 & 0.888 & 0.918 & 0.928 & 0.931 \\ 
Adjusted R$^{2}$ & 0.846 & 0.877 & 0.909 & 0.920 & 0.922 \\ 
Residual Std. Error & 0.480 & 0.429 & 0.370 & 0.346 & 0.341 \\ 
F Statistic & 68.059$^{***}$ & 82.804$^{***}$ & 98.077$^{***}$ & 118.922$^{***}$ & 106.107$^{***}$ \\ 
\hline 
\hline \\[-1.8ex] 
\textit{Note:}  & \multicolumn{5}{r}{$^{*}$p$<$0.1; $^{**}$p$<$0.05; $^{***}$p$<$0.01} \\ 
\end{longtable} 

~\\[1cm]

\begin{longtable}{@{\extracolsep{0.1pt}}lccccc} 
  \caption{Death per 100,000 (without $Y_1$ control)} 
  \label{death2} 
\small 
\\[-1.8ex]\hline 
\hline \\[-1.8ex] 
 & \multicolumn{5}{c}{\textit{Dependent variable:}} \\ 
\cline{2-6} 
\\[-1.8ex] & \multicolumn{5}{c}{death} \\ 
\\[-1.8ex] & (1) & (2) & (3) & (4) & (5)\\ 
\hline \\[-1.8ex] 
 D.14 & $-$0.073 & $-$0.070 & 0.181 & 0.198$^{*}$ & 0.231$^{*}$ \\ 
  & (0.147) & (0.149) & (0.115) & (0.112) & (0.119) \\ 
  & & & & & \\ 
 D.15 & $-$0.173 & $-$0.169 & $-$0.250$^{**}$ & $-$0.153 & $-$0.109 \\ 
  & (0.164) & (0.187) & (0.120) & (0.114) & (0.117) \\ 
  & & & & & \\ 
 D.16 & 0.016 & 0.020 & 0.131 & 0.222$^{*}$ & 0.285$^{**}$ \\ 
  & (0.166) & (0.210) & (0.132) & (0.126) & (0.129) \\ 
  & & & & & \\ 
 treat.1 & $-$0.244 & $-$0.249 & $-$0.087 & $-$0.005 & 0.020 \\ 
  & (0.158) & (0.160) & (0.114) & (0.110) & (0.113) \\ 
  & & & & & \\ 
 treat.2 & $-$0.043 & $-$0.102 & $-$0.148 & $-$0.063 & $-$0.010 \\ 
  & (0.446) & (0.485) & (0.374) & (0.351) & (0.349) \\ 
  & & & & & \\ 
 age & 0.173$^{***}$ & 0.194$^{***}$ & 0.208$^{***}$ & 0.205$^{***}$ & 0.198$^{***}$ \\ 
  & (0.025) & (0.024) & (0.016) & (0.016) & (0.016) \\ 
  & & & & & \\ 
 disability & 38.119$^{***}$ & 37.073$^{***}$ & 22.441$^{***}$ & 22.890$^{***}$ & 23.716$^{***}$ \\ 
  & (1.854) & (1.886) & (2.498) & (1.814) & (1.798) \\ 
  & & & & & \\ 
 insured & 6.994$^{***}$ & 6.908$^{***}$ & 9.375$^{***}$ & 6.682$^{***}$ & 6.185$^{***}$ \\ 
  & (1.312) & (1.333) & (1.253) & (1.288) & (1.314) \\ 
  & & & & & \\ 
 UE &  & 0.003 &  &  &  \\ 
  &  & (0.079) &  &  &  \\ 
  & & & & & \\ 
 co2 &  & 0.003$^{***}$ &  &  &  \\ 
  &  & (0.001) &  &  &  \\ 
  & & & & & \\ 
 HDI &  & 0.004 &  &  &  \\ 
  &  & (0.065) &  &  &  \\ 
  & & & & & \\ 
 obese &  &  & 5.694$^{***}$ & 5.678$^{***}$ & 4.358$^{***}$ \\ 
  &  &  & (1.458) & (1.457) & (1.196) \\ 
  & & & & & \\ 
 SSI.perc &  &  & 0.264 &  &  \\ 
  &  &  & (5.327) &  &  \\ 
  & & & & & \\ 
 crime &  &  & 0.001$^{***}$ & 0.0003 &  \\ 
  &  &  & (0.0002) & (0.0002) &  \\ 
  & & & & & \\ 
 exercise &  &  & $-$8.516$^{***}$ & $-$7.195$^{***}$ & $-$7.667$^{***}$ \\ 
  &  &  & (1.203) & (1.112) & (1.043) \\ 
  & & & & & \\ 
 hosps &  &  &  & $-$0.029 &  \\ 
  &  &  &  & (0.035) &  \\ 
  & & & & & \\ 
 hos.days &  &  &  & 0.001$^{***}$ & 0.001$^{***}$ \\ 
  &  &  &  & (0.0003) & (0.0001) \\ 
  & & & & & \\ 
 D.14:treat.1 & $-$0.097 & $-$0.096 & $-$0.217 & $-$0.177 & $-$0.174 \\ 
  & (0.219) & (0.209) & (0.150) & (0.147) & (0.151) \\ 
  & & & & & \\ 
 D.15:treat.1 & $-$0.109 & $-$0.106 & $-$0.179 & $-$0.124 & $-$0.132 \\ 
  & (0.228) & (0.226) & (0.152) & (0.146) & (0.152) \\ 
  & & & & & \\ 
 D.16:treat.1 & $-$0.108 & $-$0.104 & $-$0.242 & $-$0.178 & $-$0.177 \\ 
  & (0.226) & (0.227) & (0.159) & (0.151) & (0.155) \\ 
  & & & & & \\ 
 D.14:treat.2 & 0.057 & 0.060 & 0.070 & 0.052 & 0.052 \\ 
  & (0.589) & (0.643) & (0.505) & (0.477) & (0.471) \\ 
  & & & & & \\ 
 D.15:treat.2 & $-$0.057 & $-$0.050 & $-$0.056 & $-$0.021 & $-$0.029 \\ 
  & (0.616) & (0.674) & (0.511) & (0.479) & (0.466) \\ 
  & & & & & \\ 
 D.16:treat.2 & $-$0.117 & $-$0.110 & $-$0.074 & $-$0.041 & $-$0.025 \\ 
  & (0.639) & (0.691) & (0.581) & (0.541) & (0.521) \\ 
  & & & & & \\ 
 Constant & $-$9.188$^{***}$ & $-$10.098$^{***}$ & $-$6.252$^{***}$ & $-$5.198$^{***}$ & $-$3.899$^{***}$ \\ 
  & (1.297) & (1.275) & (1.335) & (1.248) & (1.296) \\ 
  & & & & & \\ 
\hline \\[-1.8ex] 
Observations & 196 & 196 & 196 & 196 & 196 \\ 
R$^{2}$ & 0.807 & 0.818 & 0.912 & 0.921 & 0.918 \\ 
Adjusted R$^{2}$ & 0.792 & 0.800 & 0.904 & 0.912 & 0.910 \\ 
Residual Std. Error & 0.558 & 0.547 & 0.380 & 0.362 & 0.366 \\ 
F Statistic & 54.143$^{***}$ & 46.916$^{***}$ & 102.545$^{***}$ & 107.843$^{***}$ & 117.493$^{***}$ \\ 
\hline 
\hline \\[-1.8ex] 
\textit{Note:}  & \multicolumn{5}{r}{$^{*}$p$<$0.1; $^{**}$p$<$0.05; $^{***}$p$<$0.01} \\ 
\end{longtable}

\clearpage
\appendix
\section{Health care prices}
\label{appendix: health price}
\begin{figure}[h]
\centering
\includegraphics[scale=0.6]{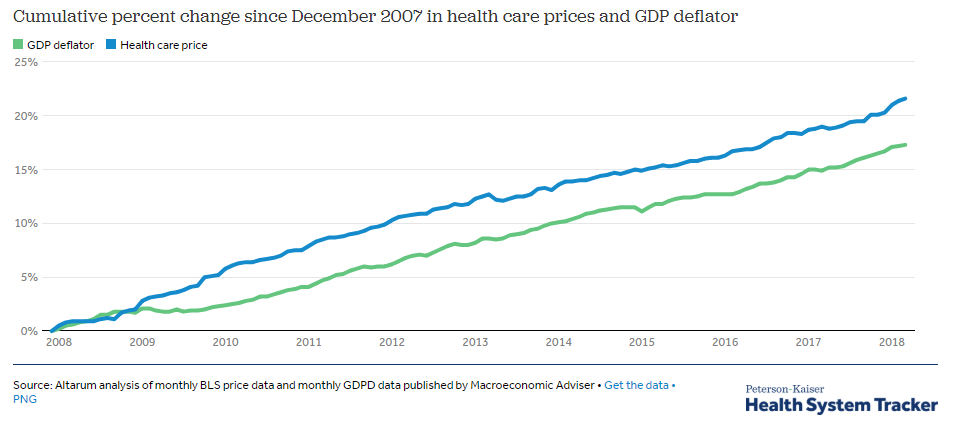}
\caption{}
\label{fig:fig 1}
\end{figure}

\section{Real Personal Consumption Expenditures: Services: Healthcare}
\label{appendix: health consumption}
\begin{figure}[h]
\centering
\includegraphics[scale=0.38]{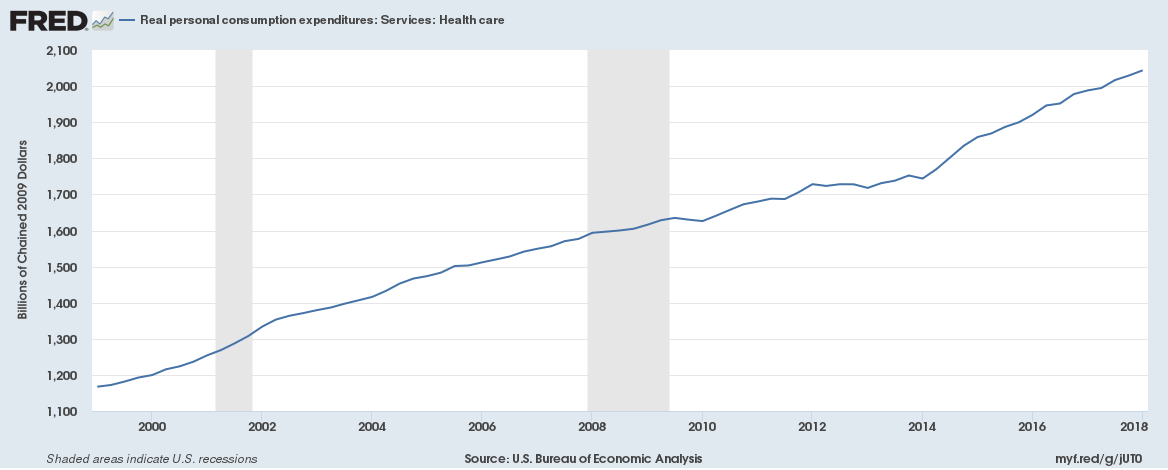}
\caption{}
\label{fig:fig 2}
\end{figure}
\citep*{a18}

~\\~\\~\\~\\

\section{Income Gap}
\label{appendix: income gap}
\begin{figure}[h]
\centering
\includegraphics[scale=1]{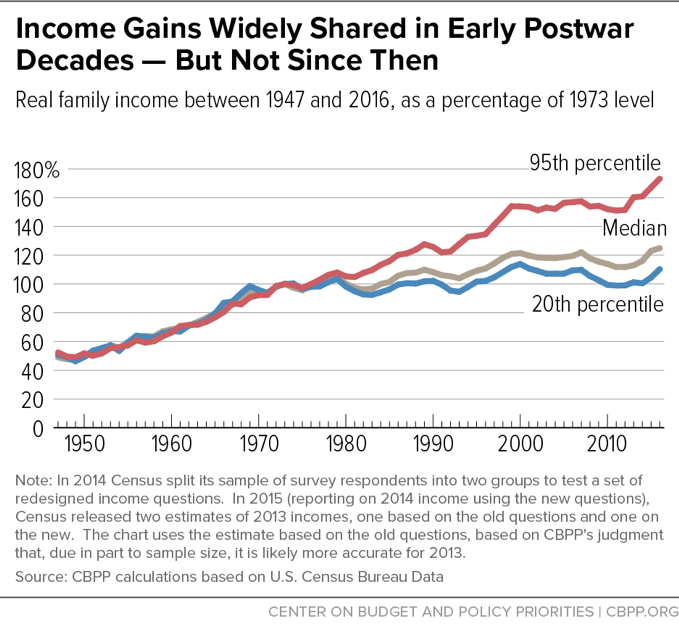}
\caption{}
\label{fig:fig 3}
\end{figure}

\newpage
\bibliographystyle{apacite}
\bibliography{ref}

\begin{thebibliography}{}

\bibitem [\protect \citeauthoryear {%
Boudreaux%
, Golberstein%
\BCBL {}\ \BBA {} McAlpine%
}{%
Boudreaux%
\ \protect \BOthers {.}}{%
{\protect \APACyear {2016}}%
}]{%
boudreaux2016long}
\APACinsertmetastar {%
boudreaux2016long}%
\begin{APACrefauthors}%
Boudreaux, M\BPBI H.%
, Golberstein, E.%
\BCBL {}\ \BBA {} McAlpine, D\BPBI D.%
\end{APACrefauthors}%
\unskip\
\newblock
\APACrefYearMonthDay{2016}{}{}.
\newblock
{\BBOQ}\APACrefatitle {The long-term impacts of Medicaid exposure in early
  childhood: Evidence from the program's origin} {The long-term impacts of
  medicaid exposure in early childhood: Evidence from the program's
  origin}.{\BBCQ}
\newblock
\APACjournalVolNumPages{Journal of health economics}{45}{}{161--175}.
\PrintBackRefs{\CurrentBib}

\bibitem [\protect \citeauthoryear {%
{Bureau of Economic Analysis}%
}{%
{Bureau of Economic Analysis}%
}{%
{\protect \APACyear {2018}}%
}]{%
gdp}
\APACinsertmetastar {%
gdp}%
\begin{APACrefauthors}%
{Bureau of Economic Analysis}.%
\end{APACrefauthors}%
\unskip\
\newblock
\APACrefYearMonthDay{2018}{}{}.
\newblock
\APACrefbtitle {Per Capita real GDP by state (chained 2009 dollars)} {Per
  capita real gdp by state (chained 2009 dollars)}\ \APACbVolEdTR{}{\BTR{}}.
\PrintBackRefs{\CurrentBib}

\bibitem [\protect \citeauthoryear {%
{Bureau of Labor Statistics, U.S. Department of Labor}%
}{%
{Bureau of Labor Statistics, U.S. Department of Labor}%
}{%
{\protect \APACyear {2017}}%
}]{%
bls}
\APACinsertmetastar {%
bls}%
\begin{APACrefauthors}%
{Bureau of Labor Statistics, U.S. Department of Labor}.%
\end{APACrefauthors}%
\unskip\
\newblock
\APACrefYearMonthDay{2017}{}{}.
\newblock
\APACrefbtitle {Local Area Unemployment Statistics} {Local area unemployment
  statistics}\ \APACbVolEdTR{}{\BTR{}}.
\newblock
\begin{APACrefURL} \url{https://www.bls.gov/lau/lastrk16.htm} \end{APACrefURL}
\PrintBackRefs{\CurrentBib}

\bibitem [\protect \citeauthoryear {%
{Center of Disease Control and Prevention}%
}{%
{Center of Disease Control and Prevention}%
}{%
{\protect \APACyear {2013}}%
}]{%
cdcheart}
\APACinsertmetastar {%
cdcheart}%
\begin{APACrefauthors}%
{Center of Disease Control and Prevention}.%
\end{APACrefauthors}%
\unskip\
\newblock
\APACrefYearMonthDay{2013}{}{}.
\newblock
\APACrefbtitle {Quickstats: Number of Deaths from 10 Leading Causes —
  National Vital Statistics System, United States, 2010} {Quickstats: Number of
  deaths from 10 leading causes — national vital statistics system, united
  states, 2010}\ \APACbVolEdTR{}{\BTR{}}.
\newblock
\begin{APACrefURL} \url{https://www.cdc.gov/mmwr/preview/mmwrhtml/mm6208a8.htm}
  \end{APACrefURL}
\PrintBackRefs{\CurrentBib}

\bibitem [\protect \citeauthoryear {%
{Federal Bureau of Investigation, Uniform Crime Reporting}%
}{%
{Federal Bureau of Investigation, Uniform Crime Reporting}%
}{%
{\protect \APACyear {2016}}%
}]{%
fbi}
\APACinsertmetastar {%
fbi}%
\begin{APACrefauthors}%
{Federal Bureau of Investigation, Uniform Crime Reporting}.%
\end{APACrefauthors}%
\unskip\
\newblock
\APACrefYearMonthDay{2016}{}{}.
\newblock
\APACrefbtitle {Crime in the United States} {Crime in the united states}\
  \APACbVolEdTR{}{\BTR{}}.
\newblock
\begin{APACrefURL}
  \url{https://ucr.fbi.gov/crime-in-the-u.s/2016/crime-in-the-u.s.-2016/topic-pages/tables/table-3}
  \end{APACrefURL}
\PrintBackRefs{\CurrentBib}

\bibitem [\protect \citeauthoryear {%
Konya%
\ \BBA {} Guisan%
}{%
Konya%
\ \BBA {} Guisan%
}{%
{\protect \APACyear {2008}}%
}]{%
konya2008does}
\APACinsertmetastar {%
konya2008does}%
\begin{APACrefauthors}%
Konya, L.%
\BCBT {}\ \BBA {} Guisan, M\BHBI C.%
\end{APACrefauthors}%
\unskip\
\newblock
\APACrefYearMonthDay{2008}{}{}.
\newblock
{\BBOQ}\APACrefatitle {What does the human development index tell us about
  convergence?} {What does the human development index tell us about
  convergence?}{\BBCQ}
\newblock

\PrintBackRefs{\CurrentBib}

\bibitem [\protect \citeauthoryear {%
{Medicaid.gov}%
}{%
{Medicaid.gov}%
}{%
{\protect \APACyear {2018}}%
}]{%
Med}
\APACinsertmetastar {%
Med}%
\begin{APACrefauthors}%
{Medicaid.gov}.%
\end{APACrefauthors}%
\unskip\
\newblock
\APACrefYearMonthDay{2018}{}{}.
\newblock
\APACrefbtitle {Medicaid} {Medicaid}\ \APACbVolEdTR{}{\BTR{}}.
\newblock
\begin{APACrefURL} \url{https://www.medicaid.gov/medicaid/index.html}
  \end{APACrefURL}
\PrintBackRefs{\CurrentBib}

\bibitem [\protect \citeauthoryear {%
{National Center for Health Statistics and others}%
}{%
{National Center for Health Statistics and others}%
}{%
{\protect \APACyear {2016}}%
}]{%
national2016healthy}
\APACinsertmetastar {%
national2016healthy}%
\begin{APACrefauthors}%
{National Center for Health Statistics and others}.%
\end{APACrefauthors}%
\unskip\
\newblock
\APACrefYearMonthDay{2016}{}{}.
\newblock
{\BBOQ}\APACrefatitle {Healthy People 2020 midcourse review} {Healthy people
  2020 midcourse review}.{\BBCQ}
\newblock
\APACjournalVolNumPages{US Department of Health and Human Services, Centers for
  Disease Control and Prevention, National Center for Health Statistics,
  Hyattsville (MD)}{}{}{}.
\PrintBackRefs{\CurrentBib}

\bibitem [\protect \citeauthoryear {%
{Social Science Research Council}%
}{%
{Social Science Research Council}%
}{%
{\protect \APACyear {2015}}%
}]{%
hdi}
\APACinsertmetastar {%
hdi}%
\begin{APACrefauthors}%
{Social Science Research Council}.%
\end{APACrefauthors}%
\unskip\
\newblock
\APACrefYearMonthDay{2015}{}{}.
\newblock
\APACrefbtitle {The Measure Of America Series: HUMAN DEVELOPMENT INDEX} {The
  measure of america series: Human development index}\ \APACbVolEdTR{}{\BTR{}}.
\PrintBackRefs{\CurrentBib}

\bibitem [\protect \citeauthoryear {%
{Social Security Administration}%
}{%
{Social Security Administration}%
}{%
{\protect \APACyear {2017}}%
}]{%
ssa}
\APACinsertmetastar {%
ssa}%
\begin{APACrefauthors}%
{Social Security Administration}.%
\end{APACrefauthors}%
\unskip\
\newblock
\APACrefYearMonthDay{2017}{}{}.
\newblock
\APACrefbtitle {Research, Statistics \& Policy Analysis} {Research, statistics
  \& policy analysis}\ \APACbVolEdTR{}{\BTR{}}.
\PrintBackRefs{\CurrentBib}

\bibitem [\protect \citeauthoryear {%
Sohn%
}{%
Sohn%
}{%
{\protect \APACyear {2017}}%
}]{%
sohn2017medicaid}
\APACinsertmetastar {%
sohn2017medicaid}%
\begin{APACrefauthors}%
Sohn, H.%
\end{APACrefauthors}%
\unskip\
\newblock
\APACrefYearMonthDay{2017}{}{}.
\newblock
{\BBOQ}\APACrefatitle {Medicaid's lasting impressions: Population health and
  insurance at birth} {Medicaid's lasting impressions: Population health and
  insurance at birth}.{\BBCQ}
\newblock
\APACjournalVolNumPages{Social Science \& Medicine}{177}{}{205--212}.
\PrintBackRefs{\CurrentBib}

\bibitem [\protect \citeauthoryear {%
Spog{\'a}rd%
\ \BBA {} James%
}{%
Spog{\'a}rd%
\ \BBA {} James%
}{%
{\protect \APACyear {1999}}%
}]{%
spogard1999governance}
\APACinsertmetastar {%
spogard1999governance}%
\begin{APACrefauthors}%
Spog{\'a}rd, R.%
\BCBT {}\ \BBA {} James, M.%
\end{APACrefauthors}%
\unskip\
\newblock
\APACrefYearMonthDay{1999}{}{}.
\newblock
{\BBOQ}\APACrefatitle {Governance and Democracy--the People’s View: A global
  opinion poll’(Gallup International Millennium Survey)} {Governance and
  democracy--the people’s view: A global opinion poll’(gallup international
  millennium survey)}.{\BBCQ}
\newblock
\APACjournalVolNumPages{Gallup International, London}{}{}{}.
\PrintBackRefs{\CurrentBib}

\bibitem [\protect \citeauthoryear {%
{State Health Access Data Assistance Center (SHADAC)}%
}{%
{State Health Access Data Assistance Center (SHADAC)}%
}{%
{\protect \APACyear {2018}}%
}]{%
shadac}
\APACinsertmetastar {%
shadac}%
\begin{APACrefauthors}%
{State Health Access Data Assistance Center (SHADAC)}.%
\end{APACrefauthors}%
\unskip\
\newblock
\APACrefYearMonthDay{2018}{}{}.
\newblock
\APACrefbtitle {State Health Compare} {State health compare}\
  \APACbVolEdTR{}{\BTR{}}.
\newblock
\begin{APACrefURL} \url{http://statehealthcompare.shadac.org/Data}
  \end{APACrefURL}
\PrintBackRefs{\CurrentBib}

\bibitem [\protect \citeauthoryear {%
{The Henry J. Kaiser Family Foundation}%
}{%
{The Henry J. Kaiser Family Foundation}%
}{%
{\protect \APACyear {2018}}%
}]{%
KFF}
\APACinsertmetastar {%
KFF}%
\begin{APACrefauthors}%
{The Henry J. Kaiser Family Foundation}.%
\end{APACrefauthors}%
\unskip\
\newblock
\APACrefYearMonthDay{2018}{}{}.
\newblock
\APACrefbtitle {Status of State Action on the Medicaid Expansion Decision}
  {Status of state action on the medicaid expansion decision}\
  \APACbVolEdTR{}{\BTR{}}.
\PrintBackRefs{\CurrentBib}

\bibitem [\protect \citeauthoryear {%
Thompson%
}{%
Thompson%
}{%
{\protect \APACyear {2017}}%
}]{%
thompson2017long}
\APACinsertmetastar {%
thompson2017long}%
\begin{APACrefauthors}%
Thompson, O.%
\end{APACrefauthors}%
\unskip\
\newblock
\APACrefYearMonthDay{2017}{}{}.
\newblock
{\BBOQ}\APACrefatitle {The long-term health impacts of Medicaid and CHIP} {The
  long-term health impacts of medicaid and chip}.{\BBCQ}
\newblock
\APACjournalVolNumPages{Journal of health economics}{51}{}{26--40}.
\PrintBackRefs{\CurrentBib}

\bibitem [\protect \citeauthoryear {%
{U. S. Bureau of Economic Analysis}%
}{%
{U. S. Bureau of Economic Analysis}%
}{%
{\protect \APACyear {2018}}%
}]{%
a18}
\APACinsertmetastar {%
a18}%
\begin{APACrefauthors}%
{U. S. Bureau of Economic Analysis}.%
\end{APACrefauthors}%
\unskip\
\newblock
\APACrefYearMonthDay{2018}{}{}.
\newblock
\APACrefbtitle {Real personal consumption expenditures: Services: Health care
  [DHLCRX1Q020SBEA]} {Real personal consumption expenditures: Services: Health
  care [dhlcrx1q020sbea]}\ \APACbVolEdTR{}{\BTR{}}.
\newblock
\begin{APACrefURL} \url{https://fred.stlouisfed.org/series/DHLCRX1Q020SBEA}
  \end{APACrefURL}
\PrintBackRefs{\CurrentBib}

\bibitem [\protect \citeauthoryear {%
{U.S. Census Bureau}%
}{%
{U.S. Census Bureau}%
}{%
{\protect \APACyear {2016}}%
}]{%
disability}
\APACinsertmetastar {%
disability}%
\begin{APACrefauthors}%
{U.S. Census Bureau}.%
\end{APACrefauthors}%
\unskip\
\newblock
\APACrefYearMonthDay{2016}{}{}.
\newblock
\APACrefbtitle {Disability Characteristics 2016 ACS 1-year estimates)}
  {Disability characteristics 2016 acs 1-year estimates)}\
  \APACbVolEdTR{}{\BTR{}}.
\PrintBackRefs{\CurrentBib}

\bibitem [\protect \citeauthoryear {%
{U.S. Energy Information Administration}%
}{%
{U.S. Energy Information Administration}%
}{%
{\protect \APACyear {2015}}%
}]{%
co2}
\APACinsertmetastar {%
co2}%
\begin{APACrefauthors}%
{U.S. Energy Information Administration}.%
\end{APACrefauthors}%
\unskip\
\newblock
\APACrefYearMonthDay{2015}{}{}.
\newblock
\APACrefbtitle {State Carbon Dioxide Emissions Data} {State carbon dioxide
  emissions data}\ \APACbVolEdTR{}{\BTR{}}.
\newblock
\begin{APACrefURL} \url{https://www.eia.gov/environment/emissions/state/}
  \end{APACrefURL}
\PrintBackRefs{\CurrentBib}

\bibitem [\protect \citeauthoryear {%
Wilson%
}{%
Wilson%
}{%
{\protect \APACyear {{\protect \bibnodate {}}}}%
}]{%
wilsoneconomic}
\APACinsertmetastar {%
wilsoneconomic}%
\begin{APACrefauthors}%
Wilson, S\BPBI E.%
\end{APACrefauthors}%
\unskip\
\newblock
\APACrefYearMonthDay{{\protect \bibnodate {}}}{}{}.
\newblock
{\BBOQ}\APACrefatitle {Economic Perspectives on Utah Medicaid Reform under the
  ACA} {Economic perspectives on utah medicaid reform under the aca}.{\BBCQ}
\newblock

\PrintBackRefs{\CurrentBib}

\bibitem [\protect \citeauthoryear {%
Yazici%
\ \BBA {} Kaestner%
}{%
Yazici%
\ \BBA {} Kaestner%
}{%
{\protect \APACyear {2000}}%
}]{%
yazici2000medicaid}
\APACinsertmetastar {%
yazici2000medicaid}%
\begin{APACrefauthors}%
Yazici, E\BPBI Y.%
\BCBT {}\ \BBA {} Kaestner, R.%
\end{APACrefauthors}%
\unskip\
\newblock
\APACrefYearMonthDay{2000}{}{}.
\newblock
{\BBOQ}\APACrefatitle {Medicaid expansions and the crowding out of private
  health insurance among children} {Medicaid expansions and the crowding out of
  private health insurance among children}.{\BBCQ}
\newblock
\APACjournalVolNumPages{Inquiry}{}{}{23--32}.
\PrintBackRefs{\CurrentBib}

\end{thebibliography}

\end{document}